\documentclass{article}
\usepackage{a4wide,generic,ds,url,times}

\InitEpsf{0.5}

\title{Discovering the Information that is lost in our Databases:\\
      {\normalsize Why bother storing data if you
       can't find the information?}}
\author{H.A. Proper and P.D. Bruza\\E.Proper@acm.org}
\date{~}

\begin{document}
   \maketitle 
{\sc Published as:}
\begin{quote}
  P.D. {Bruza} and H.A.~(Erik) {Proper}. {Discovering the Information that is lost in our Databases -- Why bother storing data if you can`t find the information?} Technical report, Distributed Systems Technology Centre, Brisbane, Queensland, Australia, 1996.
\end{quote}

   \hfill {\footnotesize \begin{minipage}{4cm}
   If you know\\[0.1cm]
   \mbox{~~~~~~}what you are looking for\\[0.1cm]
   \mbox{~~~}why are you looking\\[0.1cm]
   and if you do not know\\[0.1cm]
   \mbox{~~~~~~}what you are looking for\\[0.1cm]
   \mbox{~~~}how can you find it?\\[0.1cm]
   \\
   \mbox{~~~~~~~~}{\it Old Russion proverb}\\
\end{minipage}}

   \section{The Information Discovery Problem}

We are surrounded by an ever increasing amount of data that is stored
in a variety of databases.
In this article we will use a very liberal definition of \EM{database}.
Basically any collection of data can be regarded as a database, ranging
from the files in a directory on a disk, to ftp and web servers, 
through to relational or object-oriented databases. 

The sole reason for storing data in databases is that there is an anticipated
need for the stored data at some time in the future.
This means that providing smooth access paths by which stored information
can be retrieved is at least as important as ensuring integrity of the
stored information.
In practice, however, providing users with adequate avenues by which to
access stored information has received far less attention.

This brings us to the \EM{information discovery problem}.
In figure~\ref{\Paradigm} we have tried to portray the essential
aspects of the information discovery problem.
On one side (the right hand side), we have the information sources
as provided by the databases that are at our disposal.
These information sources, which may be aggregated into complex sources,
are characterised in some way to facilitate their discovery.
On the other side, we have a user with a certain information need.
This user is presumed to express this need in terms of an information
request.
This request will usually only be a crude description of the actual resource 
need.
Therefore, we will have to cater for further refinements of this need
as we go along.
This refinement process is usually referred to as
\EM{relevance feedback}.

We also have to take into consideration the fact that the need for 
information is there for a reason.
The need for information is born from a gap in the user's knowledge.
This gap can range from a specific need as \EM{last months
sales figures}, to the very broad \EM{relativity theory of Einstein}.
A specific need can usually be satisfied by a small collection of facts,
while a broad needs usually requires a wider variety of facts.
Observe that during the search process users may learn more and more 
about their knowledge gap, and may discover aspects of 
this gap they were initially not aware of.
This means that the actual information need of a user may change
as they are gradually exposed to new information.

Given an information request, a selection of information sources that are
considered relevant can be made.
This selection mechanism can be compared to an automatic brokering
service, matching demand to supply.
Initially, only a limited number of the selected sources can be shown to 
the user to obtain \EM{relevance feedback} from the user to further refine
the information request.
The problem of matching a given, and fixed, information request $q$ to a
set of information sources and their characterisation, corresponds to the
more traditional notion of \EM{information retrieval}.
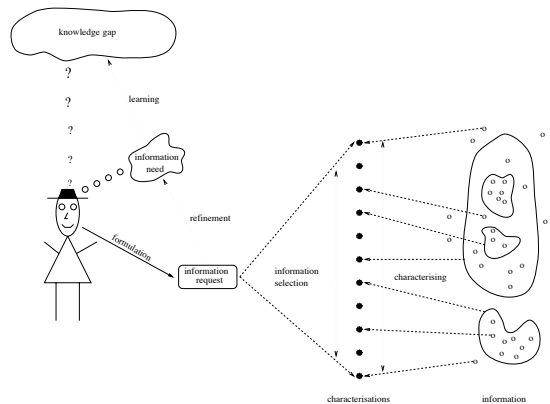
\Epsf[0.5]{\Paradigm}{Information discovery paradigm}

The information discovery problem boils down to finding a way to,
given a user's knowledge gap, find the right information sources that
will fill this gap.
Three issues play a central role here:
\begin{enumerate}
   \item formulation of information requests
   \item characterisation of information sources
   \item selection of information sources
\end{enumerate}
The formulation of information requests involves two important
aspects.
First of all, it requires some formal language in which to express
the information requests.
Secondly, obtaining a correct formulation of the \emph{true} information
need of a user is non trivial.

Good characterisation of information sources is imperative for effective
information discovery, as this fields is also subject to the old
principle of \emph{garbage-in garbage-out}.
Bad characterisations inevitably leads to the selection of irrelevant
information, or missing of relevant information.

The selection of relevant information sources for a given query $q$
is, in the mean time, a well understood problem.
For finding unstructured information sources, the research field
of information retrieval has developed a plethora of mechanisms.
However, this field is still very much in a stage characterised by lots of 
empirical testing and study.

In the remainder of this short article we will discuss some aspects
that play a role in the above issues of information discovery.
This article focuses solely on the higher level aspects of information
discovery; the conceptual level.
In \cite{Report:95:Iannella:OIL}, the HotOIL prototype is discussed in 
more technical terms.
This prototype will serve as a test bed for the ideas presented in this
article.

   \section{Formulating the Information Need}
As stated before, two aspects are involved in the formulation of 
information requests.
A language is required in terms of which requests are to be formulated,
and furthermore, obtaining a correct formulation of the \emph{true} 
information need of a user is a non trivial task.
One famous study found that sixty percent of information need 
formulations are imprecise reflections of the actual
need~\cite{Article:91:Cleverdon:CranfieldTest}.
Quite often, users have only a vague idea of the information they
indeed \emph{are} looking for, while they very well know what they are
\emph{not} looking for.

The language used for the formulation of information needs is highly 
dependent on the strategies used to help users with the formulation 
of requests.
In this section we will discuss a strategy to help users with the
formulation of their information needs, together with a language
that seems appropriate for these purposes.

\subsection{How to find what you don't know}
As mentioned before, a user's information need is born from the
existance of a gap in the knowledge of the user.
This causes an immediate problem. 
To formulate the exact information need, users must specify somehow
what their knowledge gap is, which requires them to have knowledge of
something they do not know yet.
That is after all why they are looking for it!

We therefore start out with the following simple assumption on
users:
\begin{quote} \it
   Users are able to formulate some clues about their knowledge
   gap.
\end{quote}
For example, suppose a user wants to be informed about the relation
between river polution and the migration of salmon.
This could lead to the following expressions:
\SF{polution of rivers} and \SF{migration of salmon}.
At present, we presume these expressions to be in the form of
so-called noun-phrases.

These clues originate from the user's \EM{active memory}.
The idea is now to use a strategy that allows users
to even closer approximate their actual information need.
We will try to do this by confronting the user with possible
refinements of the original clues.
For this purpose, we will have to make a further assumption on
users:
\begin{quote} \it
   Users can identify whether a clue is relevant to their
   knowledge gap.
\end{quote}
Observe that we do \EM{not} presume that a user is able to 
identify if a clue is \EM{not} relevant to their knowledge
gap.
If the system proposes a refinement of a clue that uses 
terminology that is not part of the user's current 
knowledge, then the user is not able to identify it as relevant 
or irrelevant.

With this last assumption on users, we have gained access to a user's 
passive knowledge.
In figure~\ref{\Approximate} we have depicted the process we intend to
use to approximate a user's knowledge gap.
In this figure, $i_1$, $i_2$ and $i_3$ are some initial clues 
about the user's knowledge gap, while $f_1$, $f_2$ and $f_3$ are
the more refined clues that are derived from these initial clues.
The apparant question is now, how to get from the initial clues
to the more refined clues.
Our answer to this question is \EM{query by navigation}.
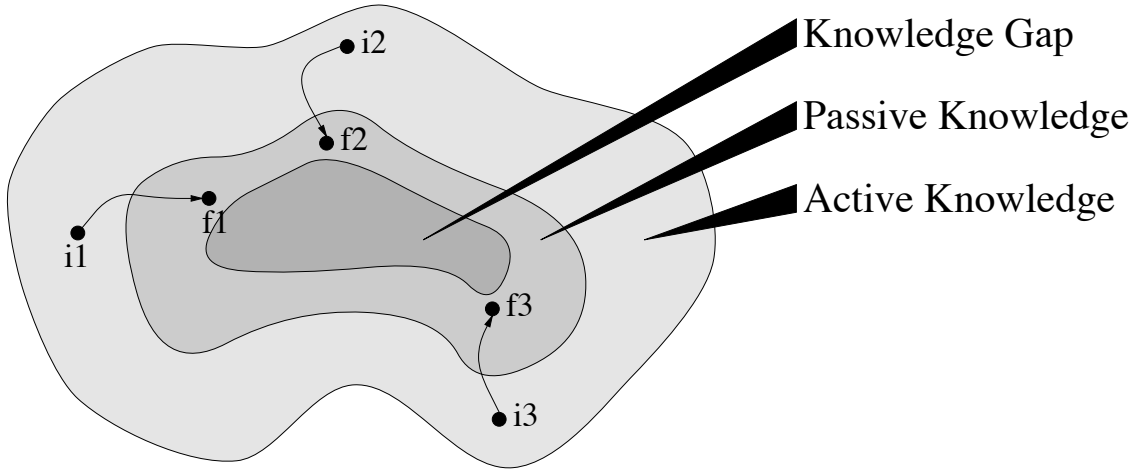
\Epsf{\Approximate}{Approximation of knowledge gap}

\subsection{Query by navigation}
We presume that for each information source we have some descriptions
of the information provided in the form of noun phrases.
For example, a gif image that depicts the proclamation by Jesus' disciples
of his resurrection could be expressed by the following noun phrase:
\[ \SF{proclamation of resurrection of Jesus by disciples} \]
This example is taken from \cite{Report:91:Bruza:InfDiscl},
where a prototype implementation of a query by 
navigation based retrieval system is discussed.
This prototype is still being used by History of Art libraries, and 
is also being sold as a commercial application.

From the above noun-phrase, called an index expression, we can derive 
part of what we call the hyperindex.
In figure~\ref{\HyperIndex} we have depicted the derived part of the
hyperindex.
This is a simple example hyperindex corresponding to a lattice, which
only deals with a breakdown of the given noun-phrase.
In reality, a hyperindex is formed by the union of
a large number of such smaller lattices, which then form a so-called
\EM{lithoid}.
Each node in the lithoid can be interpreted as a clue about the user's
knowledge gap.
Given an initial clue of this gap, the hyperindex shows us possible 
refinements (and enlargements) of this clue, allowing us to protrude into
the user's passive knowledge.

The protrusion into the user's passive knowledge, starting from
their active knowledge, is done by navigating over the hyperindex.
Hence the name \EM{query by navigation}; the information request is
formed by navigation over the hyperindex.
A sample navigation session is provided in figure~\ref{\Navigation}.
A user starts at the starting node, which contains a list of all
elementary terms from the hyperindex.
The user can then select one of these words as a first refinement.
Once a more complicated index expression has been selected,
e.g.\ \SF{resurrection of Jesus}, it becomes possible to select the
more elementary expressions that are part of the currently focussed
expression.
In the case of \SF{resurrection of Jesus} this would be
\SF{resurrection} and \SF{Jesus}.
In such a navigation session, the user basically traverses edges in
the graph of the hyperindex as shown in figure~\ref{\HyperIndex}.
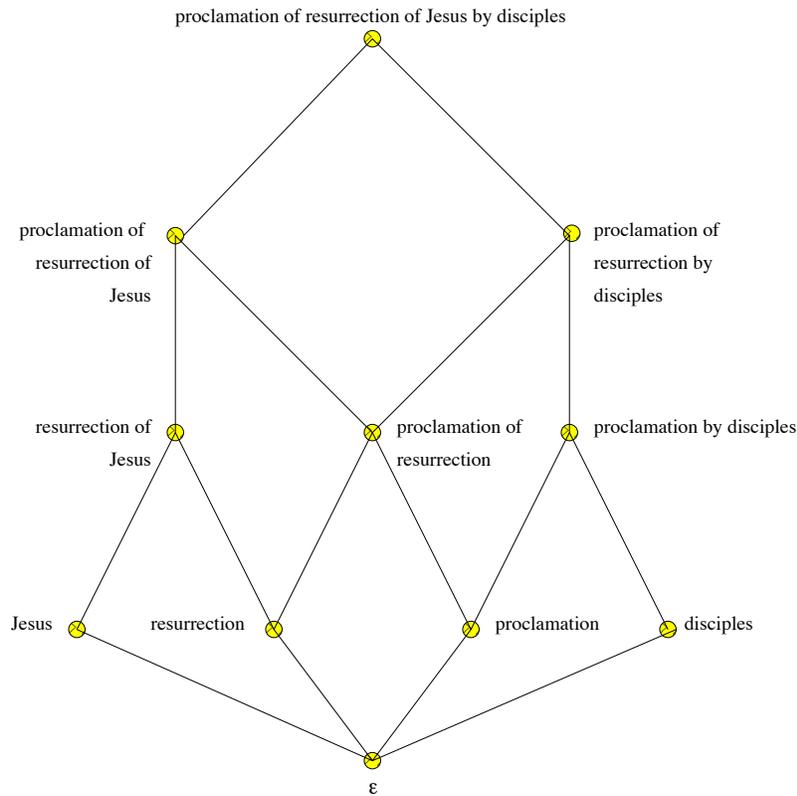
\Epsf[0.7]{\HyperIndex}{An example part of a hyperindex}

Each entry in the nodes displayed in figure~\ref{\Navigation} represents one
way to continue navigating through the hyperindex.
A node thus corresponds to a moment of choice in the search process.
The order in which the alternatives are listed in the starting node,
and nodes in general, can be based on multiple factors.
An example of such a factor is the user's past search behaviour
\cite{Report:94:Berger:IRSupport,Article:96:Bruza:NMR}.
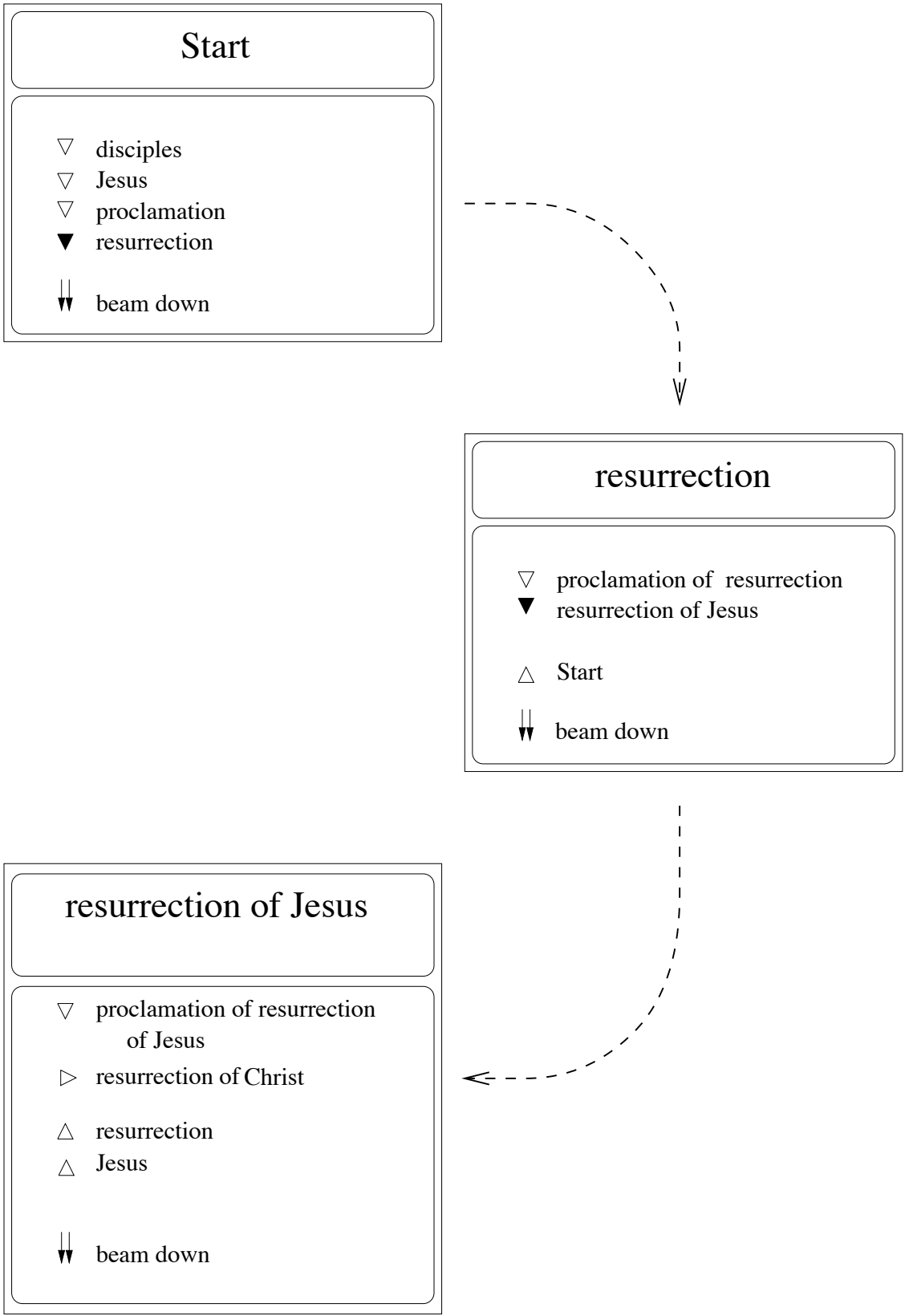
\Epsf{\Navigation}{Example navigation session}

An on-line example of a prototype information discovery tool can be
found on:
\[ {\tt http://www.dstc.edu.au/cgi-bin/RDU/hib/hib} \]
This latter prototype serves as a front-end to existing world wide 
web search tools such as Lycos and Alta-Vista.
The idea of using query by navigation has been used before
in the field of information retrieval
\cite{Article:89:Agosti:TwoLevelHyp,Article:90:Lucarella:HyperIR,
      Report:90:Bruza:IdxExp,Article:91:Agosti:TwoLevelHyp,
      Report:91:Bruza:InfDiscl,Report:91:Bruza:StratHypmed,
      Report:94:Hofstede:CSQF-QBN}.
In \cite{Report:91:Bruza:InfDiscl}, reports on empirical tests can be
found, showing the effectiveness of query by navigation. 
The use of query by navigation to support users with the formulation
of queries on structured databases has been studied in
\cite{Report:94:Hofstede:CSQF-QBN}.

\subsection{Of course I mean wave-surfing when I talk about surfing}
Whenever we as humans communicate with each other, the contextual background
is often assumed.
One way to view this background context is via a frame-based cognitive model
\cite{Book:92:Barsalou:CognitivePsych}.
The frames are constructed by attributes which may take on certain values.
For example, the attribute \SF{surfing} may take on the value \SF{wave},
thus modelling the concept \SF{wave surfing}.
It turns out that humans prime certain attributes with default values.
A mismatch in defaults between two people communicating can therefore
lead to mis-communication.

In an information discovery setting, a mis-communication between user and
discovery system may occur, usually resulting in the selection of irrelevant
information.
When we want to discover information about \SF{surfing}, while harbouring the
default \SF{wave surfing}, the system should preferably not present
information sources about \SF{internet surfing}.
An advanced information discovery system will learn a user's preferences and
anticipate further preferences based on those it has.
In \cite{Article:95:Bruza:Aboutness,Article:96:Bruza:NMR} 
strategies are discussed that allow
us to reason with user's preferences.
These strategies are based on the ideas of non-monotonic reasoning and
in particular preference logics 
\cite{Article:90:Lehman:NMR,Chapter:89:Shoham:NMR}.

A user's defaults will initially  be based on common sense.
From repositories like WordNet \cite{Article:90:Miller:WordNet}, and
Cyc \cite{Book:90:Lenat:Cyc}, we can derive default defaults.
Once a user starts navigating through a hyperindex, we can glean more
user specific defaults by observing their behaviour.
Finally, the co-occurrence of characterisations for those information 
sources that are considered relevant by the user can be used to derive
further defaults.
In \cite{Article:96:Bruza:NMR} some strategies to derive these
defaults have been discussed.

   \section{Characterisation of Information Sources}
Effective information discovery starts with good characterisations of
information sources.
The old principle of \emph{garbage-in garbage-out} also applies to
information discovery.
Bad characterisations inevitably leads to the selection of irrelevant
information, or missing of relevant information.

The characterisation of information sources involves two crucial aspects.
Again, an adequate language is needed in which the
characterisation of information sources can be expressed.
An intersting question is of course whether this language should be
the same as the language in which information requests can be expressed. 
The second aspect, the actual characterisation of information
sources, is absolutely crucial.
A complicating factor in the characterisation process is the
wide variety and sheer volume of available sources;
making manual characterisation impractical.
Resource characterisation also raises questions like:
\EM{who should do it}, \EM{where should it be done}, and \EM{when}.
In the context of the net, we cannot simply presume that information
providers also provide characterisations.
Even when information providers do characterise their sources, we are at
their mercy with regards to the quality, protocols, and languages used.

Deriving characterisations from information sources depends
very much on the medium and purpose of the source.
For texts, automatic mechanisms to derive
characterisations exist
\cite{Article:77:Maron:InfRetr,Article:77:Hutchinson:Aboutness,
      Article:91:Ruge:NLTextProcessing,Article:93:Rama:TextGrammar}.
However, the more effective of these techniques rely on statistics that are
generated from the given universe of documents.
In the context of the net, these statistics are hard to obtain due to the 
openess of the net.
Furthermore, the semantic information stored with information
sources is typically non-existent~\cite{Article:95:Lynch:ResourceDiscovery}.
As a consequence, characterisation of information sources is a very difficult
problem which is further compounded by aggregation of information sources.

Currently, the characterisation of images and sound is far from
automatable.
This means that unless these sources are explicitly characterised,
or implicitly by being embedded in e.g.\ a web document, they
can not be found by a search tool.
An interesting idea on how to characterise non-textual information sources is
discussed in~\cite{Article:93:Dunlop:ImageCharacterisation}.
There it is suggested to use documents that contain, or refer to,
the non-textual information source as a base for its characterisation.

For characterisation of databases, it is also very important
to make a distinction between databases that have an underlying
\emph{conceptual} schema and those that do not.
A conceptual schema provides a semantically rich description
of the structure of the stored data
\cite{Book:92:Batini:ConceptualDBDesign,Book:94:Elmasri:DBFundamentals,
      Book:94:Halpin:ORM}.
This description provides valuable information that can help
determine how relevant a given database is for an information
need.
Unfortunately, however, most legacy systems do not have an associated
conceptual schema.
This is even more unfortunate when we realise that most existing
databases on the net fall in this category.
This means that effective characterisation of such databases is difficult
due to the dearth of semantic information about the contents.

We are presently looking for a characterisation language that will allow
us to characterise a wide range of types of information sources.
Our present thoughts are to start from index expressions, basically noun 
phrases, as defined in \cite{Article:78:Craven:IndexExpressie,
Report:90:Bruza:IdxExp}, and apply linguistic principles to obtain a
a linguistically normalised format \cite{Book:95:Allen:NatLanguage}.
Using this latter normalisation, index expressions like:
\[ \SF{success of tourism in Australia} \]
and
\[ \SF{how tourism in Australia succeeds} \]
would map to the same \EM{logical} representation.
An interesting challenge is to be able to deal with multiple languages.

   \section{Selection of Information sources}
The selection of relevant information sources for a given query $q$
is, in the mean time, a well understood problem.
For finding unstructured information sources, the research field
of information retrieval has developed a myriad of mechanisms.
However, this field is still very much in a stage characterised by lots of 
emperical testing and study.
A well-defined theoretical account of the underlying matching mechanisms
does not exist yet 
\cite{Article:86:Rijsbergen:IRLogic,Article:86:Rijsbergen:IRFramework,
      Article:89:Rijsbergen:IRLogic}.

In the context of structured information sources, like information
stored in relational databases, the selection boils down to answering 
a query that is formulated in some language like SQL.
Note: from our point of view, each object stored in a relational or
object-oriented database is an information source.
As structured databases focus on structured data only the matching mechanisms
used are well understood and relatively simple. 
Given a query $q$, the result is known exactly, and recall and precision
are always 100\%.
Observe that even though these structured databases
have received most commercial interest so-far, the majority of stored 
information is \EM{not} contained in this kind of databases.
The majority of information is actually stored in the form of textual
documents \cite{Article:90:Wiggens:95rule}, and quite possibly not even 
in an electronic format.

Most of the current information source selection mechanisms do not cater
for elevance feedback or cognitive feedback.
They simply presume that the user was able to come up with an exact definition
of their information need.
As argued above, we consider this to be an unrealistic and unpractical 
assumption.
It certainly makes the life of query mechanisms easier, but does not help the
users of these systems.
We therefore propose the use of selection mechanisms that are more
attuned to user preferences \cite{Article:96:Bruza:NMR}.

   \section{Discussion}

The critical questions one may ask about the above presented
ideas is whether they will work in practice.
With regards to the formulation of information needs, 
empirical tests as can be found in \cite{Report:91:Bruza:InfDiscl}
lead us to believe that the use of query by navigation will help users
better find the information they are \EM{really} looking for.
At present, the Resource Discovery Unit is planning further empirical
tests to verify the effectiveness of query by navigation in the
context of searching on the world wide web.

For the selection of information sources, the use and practicality
of user preferences and defaults still needs to be tested.
A first prototype of a \EM{preference reasonor} has been developed
in Prolog, and the next step will be to integrate this with the
HotOIL \cite{Report:95:Iannella:OIL} prototype, and use this prototype 
as a base for evaluation and experimentation.

The characterisation of information sources needs further investigation.
Open issues are the automatic derivation of characterisations from
textual sources, and in particular the semantic normalisation of
resulting index expressions as discussed earlier.
In this area we will cooperate with the Software Engineering and
Linguistics Departments from the University of Nijmegen, which
have extensive experience in the development of parsers and lexica
for natural languages.
Also, the characterisation of legacy databases needs attention.
For databases with a proper conceptual schema, the verbalisations in
the conceptual schema can be used as a base for characterisations.
In the case of legacy databases, a `quick and dirty' reverse engineering
step seems to be unavoidable.
Finally, characterisation of non-textual information sources, like graphics,
video and audio, is still very much an open field.

Finally, what may initially sound less relevant from a practical point of view, 
but which will have a significant impact on the development of information
discovery theories, is the development of an underlying theory of information.
We talk and think about information retrieval and information discovery systems
without paying much attention to the question \EM{what is information}.
We are currently looking at the work done in e.g.\ situation theory
\cite{Article:90:Barwise:SitTheory,Book:89:Barwise:SitTheory,
      Book:91:Devlin:SitTheory}, and information theory 
\cite{Book:86:Landman:SitTheory}, to develop such an underlying theory
for information discovery.

   \Biblio{alpha}{all}
\end{document}